\documentclass[12pt]{article}
\usepackage{graphicx}
\begin{document}
\global\arraycolsep=2pt 
%
%
%
\thispagestyle{empty} 
\begin{titlepage}    
\begin{flushright}
TNCT-0901
\end{flushright}
\vspace{1cm}
\begin{center}
  {\Large{\bf 
   Influence of current mass on the spatially 
     inhomogeneous chiral condensate} }
\end{center}                                   
\vspace{0.8cm}
              
\begin{center}
Shinji Maedan            
   \footnote{ E-mail: maedan@tokyo-ct.ac.jp}   
           \\
\vspace{0.8cm}
{\sl  Department of Physics, Tokyo National College of Technology,
        Kunugida-machi, Hachioji, Tokyo 193-0997, Japan
                                   }
\end{center}                                               
            
\vspace{0.5cm}
              
\begin{abstract}  
\noindent       
It is known that, in the chiral limit, spatially inhomogeneous chiral condensate
  occurs in the  Nambu-Jona-Lasinio (NJL) model at finite density within a 
  mean-field approximation.
We study here how an introduction of current quark mass affects the ground
  state with  the spatially inhomogeneous chiral condensate.
Numerical calculations show that, even if the current quark mass is introduced,
  the spatially inhomogeneous chiral condensate can take place.
In order to obtain the ground state, the thermodynamic potential is
  calculated with a mean-field approximation.
The influence of finite current mass on the thermodynamic potential consists
  of following two parts.
One is a part coming from the field energy of the condensate, which favors
  inhomogeneous chiral condensate.
The other is a part coming from the Dirac sea and the Fermi sea, which favors
  homogeneous chiral condensate.
We also find that when the spatially inhomogeneous chiral condensate
  occurs, the baryon number density becomes spatially inhomogeneous.
\end{abstract} 
\vskip 1.5cm
\begin{center}
%
%
%
%
\end{center}
\vfill            
\end{titlepage}
%
%
%
\setcounter{page}{1}
\section{Introduction} 

Quantum chromodynamics (QCD) at finite density is one of interesting
  topics in these days.
The study of this field will help us to understand the physics of neutron
  stars, compact stars, and heavy ion collisions.
In the vacuum state of QCD (quark chemical potential $\mu =0$) at low
  temperature, the chiral symmetry is broken
  spontaneously and the confinement occurs.
On the other hand, at extremely high densities ($\mu$ has very large values)
  and low temperature, color superconductivity will be realized
  in QCD \cite{rf:Bar,rf:BaiLov}.
At such high density region, the perturbative calculation with the gauge 
  coupling $g$ is possible due to asymptotic freedom.

Now, at the moderate density region (moderate value of $\mu$) where
  the coupling $g$ is not small, how a ground state of QCD becomes?
The method of perturbation of the coupling $g$ cannot be used, and also it is
  difficult to apply the lattice QCD simulations to a system with finite density.
Therefore, people have used effective theories of QCD such as the 
   Nambu-Jona-Lasinio (NJL) model \cite{rf:NamJon,rf:Kle} to study the physics
  at the moderate density region\cite{rf:Bub}.
One of the interesting topics of these researches is that spatially
  inhomogeneous chiral condensate occurs in the ground state at
  the moderate density \cite{rf:TatNak,rf:NakTat,rf:Nic1,rf:Nic2}. 

Before discussing this topic in detail, let us look back the idea that fermionic
  condensate becomes spatially inhomogeneous at finite density.
The possibility of spatially inhomogeneous chiral condensate in QCD at
  finite density was first discussed in Ref\cite{rf:DerGriRub}.
The authors in Ref\cite{rf:DerGriRub}
  have shown that in the limit, $N_c$ (number of colors) $ \rightarrow \infty $,
  the spatially inhomogeneous chiral condensate occurs in the ground state
  of QCD at extremely high density region such that the relation 
  $ g^2 N_c \ll 1$ holds.
That state is the standing wave ground state having the wave number $2\mu$,
  in which particle and hole with the same Fermi momentum 
  $ \mbox{\boldmath  $p$} \, (\vert  \mbox{\boldmath  $p$} \vert = \mu) $
  condense (Overhauser effect \cite{rf:Ove}).
When one takes the number of colors $N_c$ of QCD to the realistic number three,
  however, it is shown by the later researches that the BCS effect 
  (particle-particle condensation) is superior to the Overhauser effect
  (particle-hole condensation) 
 and the color superconductor is realized at
  extremely high density \cite{rf:ShuSon,rf:ParRhoWirZah,rf:RapShuZah}.
In some two-dimensional models, it is argued that spatially inhomogeneous
 chiral condensate also occurs at finite density.
In the chiral Gross-Neveu model with the limit $N \rightarrow \infty $ at
  finite density, it is shown that both scalar $  \langle \bar\psi\psi  \rangle $
  and pseudoscalar 
  $  \langle \bar\psi  i \, \gamma_5 \,  \psi  \rangle $ 
  become spatially inhomogeneous in the ground state
   \footnote{
  In the chiral Gross-Neveu model, finite
  bare quark mass case has also been studied  \cite{rf:SchThi}.
  When the bare quark mass is taken to be finite, the chiral angle depends
  on the value of the bare mass and the baryon density becomes 
  spatially inhomogeneous.} \cite{rf:SchThi,rf:Ohw}.

Now, let us return to the topic mentioned above.
In Ref \cite{rf:TatNak,rf:NakTat}, the NJL model with $N_c = 3$ and 
  $ N_f$ (number of flavors) $=2$ is considered at moderate density.
Assuming the following mean-field,
\begin{equation}
   \langle \bar\psi\psi  \rangle 
   = \triangle \cdot  \cos \,(  \mbox{\boldmath  $q$}   \cdot
                   \mbox{\boldmath  $r$}  ), \hskip 1cm
  \langle \bar\psi i \, \gamma_5 \, \tau_3 \, \psi \rangle
   = \triangle \cdot \sin \,( \mbox{\boldmath  $q$}
            \cdot  \mbox{\boldmath  $r$} ),
  \label{aa}
\end{equation}
where $ \mbox{\boldmath  $q$} $ is a wave number vector, the authors in
  Ref \cite{rf:TatNak,rf:NakTat} obtain the ground state of that model by
  finding the minimum value of the thermodynamic potential in the 
  mean-field approximation.
In the chiral limit, they find numerically that the ground state has non-zero
  value of $ \mbox{\boldmath  $q$} $ at low temperature and a high density
  region, namely spatially inhomogeneous dual chiral
  condensate, Eq.(\ref{aa}), occurs at that density region.
Furthermore, the quark number density $ \langle \psi^\dagger \psi  \rangle $
  becomes spatially homogeneous in the chiral limit.

In Ref \cite{rf:Nic2}, the $T-\mu$ phase diagram of the NJL-type model is studied
   around the chiral critical point in the chiral limit, which includes spatially
   inhomogeneous chiral condensate.
There, the condensate is restricted to only scalar form instead of allowing
   both scalar and pseudoscalar form like  Eq.(\ref{aa}).
With more general ansatz for the scalar condensate, two phase transitions
   are found \cite{rf:Nic2}.
One is from the homogeneous massive phase to inhomogeneous phase,
   and the other is from the inhomogeneous phase to chirally symmetric phase,
   both of which are the second order.
In the real world, however, the current quark mass is not zero and the 
  pion (NG boson) has a mass of about $135 {\rm MeV}$.
Recently, the same issue is studied for the finite current
   quark mass \cite{rf:Nic1}.
In general, the thermodynamic potential $\Omega$ is a function of the complex
   order parameter $M({\bf r})$,
\begin{equation}
  M( {\bf r} ) = m- 2 G_s \left\{  \langle \bar\psi\psi  \rangle 
    + i  \langle \bar\psi  i \, \gamma_5 \, \tau_3 \,  \psi  \rangle  \right\},
  \label{ab}
\end{equation}
and the stationary constraint, 
   $ \delta \Omega / \delta M ( {\bf r} )^* =0 $,
   is held in the ground state.
In Ref \cite{rf:Nic1}, the stationary constraint is solved when the order
   parameter $M({\bf r})$ is restricted to real
   (i.e., pseudoscalar condensate = 0).
The numerical calculation using that real solution shows that the
   qualitative feature of the phase diagram remains unchanged.
Now, how about the case where 
   the order parameter $M({\bf r})$ 
   takes a complex value
   (i.e., both scalar and pseudoscalar condensate)?
Unfortunately, it is very difficult to solve the stationary constraint with
   finite current mass when $M({\bf r})$ is a complex value.
Although no solution is known, we need to study 
    the case where the order parameter $M({\bf r})$ 
    is complex 
    by, for example, assuming
    ansatz for $M({\bf r})$.

In this paper,
   we study the massive NJL model in the case where  the order parameter
    is complex, that is, both scalar and pseudoscalar condensate exist.
Concretely,
  we study how an introduction of finite current mass affects
  the ground state with the spatially inhomogeneous
  dual chiral condensate, Eq.(\ref{aa}), in Ref  \cite{rf:TatNak,rf:NakTat}.
Does the dual chiral condensate remain spatially inhomogeneous?
And, does the quark number density remain spatially homogeneous?
To find out these problems at moderate baryon density, we use the NJL
  model with finite current quark mass \cite{rf:Nic1} (not the chiral limit case as
  in Ref \cite{rf:TatNak,rf:NakTat}).
Here, let us compare the ansatz, Eq.(\ref{aa}), and that in Ref \cite{rf:Nic2}.
The ansatz, Eq.(\ref{aa}), allows the pseudoscalar condensate to occur in
   addition to the scalar condensate, while the ansatz in Ref  \cite{rf:Nic2}
   restricts the condensate to the scalar form.
In this respect, we can say that Eq.(\ref{aa}) is more general than the ansatz
   in  Ref \cite{rf:Nic2}.
On the other hand, concerning Fourier analysis of the assumed condensate,
    Eq.(\ref{aa}) is restricted to monochromatic wave, while  the ansatz in Ref  \cite{rf:Nic2}
   contains various higher harmonics.
From the viewpoint of this, we can say that the  ansatz in Ref  \cite{rf:Nic2} is
   more general than  Eq.(\ref{aa}).
Now, we return to the discussion of the NJL model.
In the chiral limit, the thermodynamic potential can be calculated
  \cite{rf:TatNak,rf:NakTat} in a
  mean-field approximation under the assumption Eq.(\ref{aa}).
On the other hand, when the current mass is taken to be finite, it is
  difficult to calculate the thermodynamic potential because the fermion
  propagator depends on the space coordinates $\mbox{\boldmath  $r$}$
  explicitly.
In order to avoid this difficulty, we will expand in powers of the
  current mass $m$.
The thermodynamic potential $\omega$ will be calculated up to first order
  $O(m)$ and the ground state is obtained by finding a minimum value of
  $\omega$.

This paper is organized as follows.
The NJL model with $N_c =3$ and $N_f =2$ is introduced in section 2.
We assume that mean fields 
  $  \langle \bar\psi\psi  \rangle $ and
  $  \langle \bar\psi  i \, \gamma_5 \, \tau_3 \,  \psi  \rangle $,
  Eq.(\ref{aa}), exist at finite density, and the thermodynamic potential 
  is calculated analytically in a mean-field approximation.
In section 3, the numerical calculations of the thermodynamic potential 
  at zero temperature are carried out and we obtain the ground state
  which minimize the thermodynamic potential.
For a given quark chemical potential $\mu$, these numerical calculations enable
  us to find whether the spatially inhomogeneous chiral condensate is
  realized ($ \mbox{\boldmath  $q$} \ne 0 $)
  or not  ($ \mbox{\boldmath  $q$} = 0 $).
In section 4, we show that the quark number density becomes spatially
  inhomogeneous when the spatially inhomogeneous chiral condensate 
  is realized in the case of finite current quark mass.
Section 5 is devoted to conclusions.
%
%
%
\section{The Nambu-Jona-Lasinio model and \\ standing wave ansatz}

The Lagrangian of the NJL model is
\begin{equation}
  {\cal L} = \bar\psi(i \partial \!\!\!/ -m)\psi 
  + G \left[(\bar\psi\psi)^2+(\bar\psi i \, \gamma_5 \, {\vec \tau} \, \psi)^2\right],
  \label{ba}
\end{equation}
where the number of colors is $N_c =3$ and the number of flavors is 
  $N_f =2$.
Note that the current quark mass is taken to be finite and we set
  $m_u =m_d \equiv m >0$.
Now, we suppose the following mean fields exist at finite density
  \cite{rf:TatNak,rf:NakTat},
\begin{equation}
   \langle \bar\psi\psi  \rangle = C \cos \,( \mbox{\boldmath  $q$}
                       \cdot  \mbox{\boldmath  $r$} ), \hskip 1cm
  \langle \bar\psi i \, \gamma_5 \, \tau_3 \, \psi \rangle = C \sin \,
            ( \mbox{\boldmath  $q$} \cdot \mbox{\boldmath  $r$}  ),
  \hskip 1cm   ( C < 0 ),
  \label{bb}
\end{equation}
where $ \mbox{\boldmath  $q$} $ is the wave number vector and 
  $C$ is a constant.
The other components are assumed to vanish,
  $  \langle \bar\psi i \, \gamma_5 \, \tau_1 \, \psi \rangle 
     =  \langle \bar\psi i \, \gamma_5 \, \tau_2 \, \psi \rangle =0$.
When  $ \mbox{\boldmath  $q$} $ vanishes, these have the usual forms
   $ \langle \bar\psi\psi  \rangle = C $ and
   $ \langle \bar\psi i \, \gamma_5 \, \tau_3 \, \psi \rangle = 0$.
These mean fields are put on the chiral circle,
  $ \langle \bar\psi\psi \rangle^2 + \langle \bar\psi i \, \gamma_5 \, \tau_3 \, \psi \rangle^2  = C^2 $.

Here we shall make a comment on the ansatz Eq.(\ref{bb}).
If the chiral limit, $m \rightarrow 0$, is considered, the bottom of
  the effective potential is
  the chiral circle, and it would be natural for 
  $  \langle \bar\psi\psi  \rangle $ and
  $  \langle \bar\psi  i \, \gamma_5 \, \tau_3 \,  \psi  \rangle $ 
  to be put on that chiral circle.
However, when the current mass $m$ is finite,  the effective potential 
  slightly tilts in the direction of  $  \bar\psi \psi $ and
  the shape of the effective potential becomes very complicated.
In such case, there might exists another ansatz which has more
  appropriate function of $\mbox{\boldmath  $r$}$ than the
  trigonometrical function used in Eq.(\ref{bb}).
This point will be discussed in subsection 3.1 and section 5.
In the present paper, we use the ansatz Eq.(\ref{bb}) for simplicity.

One can choose the direction of the wave number vector
  $\mbox{\boldmath  $q$} $  as
  $ \mbox{\boldmath  $q$} =(0,0,q), ( q \geq 0 )$
  without loss of generality.
The mean-field approximated Lagrangian becomes 
\begin{equation}
   {\cal L_{\rm MF} }
  =  \bar\psi \left[ i \partial \!\!\!/ -m + \mu \, \gamma_0
    - M \exp \left( i \gamma_5 \, \tau_3  \mbox{\boldmath  $q$}  \cdot {\bf r} 
                        \right) \right] \psi
   -\frac{M^2}{4 G},
  \label{bd}
\end{equation}
where the parameter $M$ has been defined as
\begin{equation}
  M \equiv -2 G C = 2 G \sqrt{
    \langle \bar\psi\psi \rangle^2
        + \langle \bar\psi i \, \gamma_5 \, \tau_3 \, \psi \rangle^2 },
  \label{bc}
\end{equation}
and $\mu$ is the quark chemical potential.
In the path integral representation, the bilinear form of fermion in
  ${\cal L_{\rm MF} } $ can be written as
\begin{eqnarray}
  & &  \int {\cal D} \, \bar\psi  {\cal D} \psi \exp i \int d^4 x \,  
             \bar\psi \left[ i \partial \!\!\!/ -m + \mu \, \gamma_0
               - M \exp \left( i \gamma_5 \, \tau_3  \mbox{\boldmath  $q$} 
                  \cdot \mbox{\boldmath  $r$}  \right) \right] \psi      \nonumber  \\
  &=& \int {\cal D} \, \bar\psi'  {\cal D} \psi' \exp i \int d^4 x \,  
             \bar\psi' \left[ i \gamma^\mu ( \partial_\mu + {i \over 2} \gamma_5 \, \tau_3 q_\mu )
              -M + \mu \, \gamma_0
               - m \exp \left( - i \gamma_5 \, \tau_3  \mbox{\boldmath  $q$}  
                       \cdot  \mbox{\boldmath  $r$}  \right) \right] \psi'   \nonumber  \\
  &=& \int {\cal D} \, \bar\psi'  {\cal D} \psi'                                             \nonumber  \\
  & &  \times  \exp i \int d^4 x \,  
             \bar\psi' \left[ i \gamma^\mu ( \partial_\mu + {i \over 2} \gamma_5 \, \tau_3 q_\mu )
              -M_t + \mu \, \gamma_0
               - m \left\{  \exp \left( - i \gamma_5 \, \tau_3  \mbox{\boldmath  $q$} 
                      \cdot  \mbox{\boldmath  $r$}  \right) -1
                     \right\} \right] \psi',                     \nonumber              \\
  \label{be}
\end{eqnarray}
where  $ \psi' \equiv \exp \left\{ + {i \over 2} \, \gamma_5 \, \tau_3  \mbox{\boldmath  $q$} 
                 \cdot  \mbox{\boldmath  $r$}  \right\} \psi $
   and $ M+m \equiv M_t \geq m $. 
The propagator $S$ of the field $\psi' $ is
\begin{equation}
  S = \frac{1}{  i \gamma^\mu ( \partial_\mu + {i \over 2} \gamma_5 \, \tau_3 q_\mu )
              -M_t + \mu \, \gamma_0
               - m \left\{  \exp \left( - i \gamma_5 \, \tau_3  \mbox{\boldmath  $q$} 
                      \cdot  \mbox{\boldmath  $r$} \right) -1 \right\}  }.
  \label{bf}
\end{equation}
If the wave number vector $\mbox{\boldmath  $q$}$ vanishes, $S$ represents
  a propagator of free fermion with a mass $M_t$.

When the wave number vector $\mbox{\boldmath  $q$} \neq 0 $ and
  the current quark mass $m \neq 0$, the denominator of the propagator $S$
  depends on the space coordinates $\mbox{\boldmath  $r$}$ explicitly,
  so that it becomes difficult to calculate the expression involving $S$.
To avoid this difficulty, we separate $S^{-1}$ into two,
\begin{equation}
      S^{-1} = S_0^{-1} - V_m,
  \label{bg}
\end{equation}
where
\begin{eqnarray}
  S_0^{-1}  & \equiv &   i \gamma^\mu ( \partial_\mu + {i \over 2} \gamma_5 \, \tau_3 q_\mu )
              -M_t + \mu \, \gamma_0,        \nonumber  \\
  V_m & \equiv &  m \left\{  \exp \left( - i \gamma_5 \, \tau_3  \mbox{\boldmath  $q$} 
                 \cdot  \mbox{\boldmath  $r$}  \right) -1   \right\}.
  \label{bh}
\end{eqnarray}
Using the identity 
\begin{equation}
  {1 \over A}-{1 \over B} = {1 \over B} (B-A) {1 \over A},
  \label{bi}
\end{equation}
for any noncommutable operators $A$ and $B$, we obtain
\begin{equation}
  S = S_0 + S_0 V_m S.
  \label{bj}
\end{equation}
From this equation, we have
\begin{equation}
  S = S_0 + S_0 V_m S_0 + S_0 V_m S_0 V_m S_0 + \cdots.
  \label{bk}
\end{equation}
Here, it should be noted that in the chiral limit, $ m \rightarrow 0$,
   we have, $ V_m  \rightarrow 0$. 
Since the current quark mass $m$ is the smallest energy scale in the system, 
  we treat the $V_m \sim O(m) $ as perturbative part and neglect higher
  order $O( V_m^2 )$. 
We shall use the approximate equation (\ref{bk}) to calculate the expression
  involving $S$, as will be done in section 4.
For the expression involving $S^{-1}$, the separation Eq.(\ref{bg}) will be used as follows.

The thermodynamic potential $\omega$ is useful so as to find the ground
  state of the system with finite density $\mu \ge 0$ and finite 
  temperature $ T \ge 0$.
In the imaginary time formulation \cite{rf:Kap}, $\tau =i t$, the
  mean-field approximated thermodynamic potential $\omega$
  has the form,
\begin{eqnarray}
  \omega 
  &=&  -{T \over V} \ln N' \int {\cal D}  \, \bar\psi  {\cal D} \psi \exp  \int_{0}^{\beta} d \tau
                     \int d^3 x \, {\cal L}_{\rm MF}               \nonumber  \\
  &=&  \frac{M^2}{4 G} 
          -{T \over V}  \, \ln  \int {\cal D}  \, \bar\psi'  {\cal D} \psi' \exp  \int_{0}^{\beta} d \tau
                     \int d^3 x \,  \bar\psi'  S^{-1} \psi'.
  \label{bl}
\end{eqnarray}
Here $ V=L^3$ is the volume, $\beta$ is $1/T$, and the irrelevant constant is omitted.
Since the path integral calculation in Eq.(\ref{bl}) is difficult, we regard 
  the term $V_m$ in Eq.(\ref{bg}) as perturbative part and obtain $\omega$
  up to the order $O(V_m)$.
We separate the action 
  $   \int_{0}^{\beta} d \tau   \int d^3 x \,  \bar\psi'  S^{-1} \psi' \equiv A $
  into two parts referring to Eq.(\ref{bg}),
\begin{equation}
  A = A_0 + A_m.
  \label{bm}
\end{equation}
Here, $A_0$ and $A_m$ are
\begin{eqnarray}
  A_0  & \equiv &  \int_{0}^{\beta} d \tau   \int d^3 x \,  \bar\psi'  S_0^{-1} \psi',   \nonumber \\
  A_m & \equiv &  \int_{0}^{\beta} d \tau   \int d^3 x \,  \bar\psi'  ( -V_m \, ) \psi',
  \label{bn}
\end{eqnarray}
and note that  in the chiral limit, $ m \rightarrow 0$,
   one has, $ A_m  \rightarrow 0$. 
Expanding $ \exp (A_m)$, we have
\begin{eqnarray}
  w   &=&  \frac{M^2}{4 G} 
          -{T \over V} \, \ln  \int {\cal D}  \, \bar\psi'  {\cal D} \psi' \exp A  \nonumber \\
  & = &   \frac{M^2}{4 G} 
          -{T \over V}  \, \ln  \int {\cal D}  \, \bar\psi'  {\cal D} \psi' \exp ( A_0 )
           \sum_{l=0}^{\infty} \, { 1 \over l ! } \, A_m^l             \nonumber \\
  & = &   \frac{M^2}{4 G} 
            -{T \over V}  \,  \left[  \ln  \int {\cal D}  \, \bar\psi'  {\cal D} \psi' \exp ( A_0 )
             + \frac{ \int {\cal D}  \, \bar\psi'  {\cal D} \psi'  A_m \exp ( A_0 ) }
                        { \int {\cal D}  \, \bar\psi'  {\cal D} \psi'          \exp ( A_0 ) }
            + O(V_m^2 \,)  \right].
  \label{bo}
\end{eqnarray}
The leading term in Eq.(\ref{bo}) is
\begin{eqnarray}
    -{T \over V}  \,  \ln  \int {\cal D}  \, \bar\psi'  {\cal D} \psi' \exp ( A_0 )
  & = &   -{T \over V} \, N_c  \ln  {\rm Det}  \left( \frac{\, S_0^{-1}}{T} \right)           \nonumber \\
   &=&   - T N_c \int  \frac{ d^3 p}{ (2 \pi)^3 } \sum_{j=-\infty}^{\infty}
                 \ln \det  \left( \frac{\, S_0^{-1}}{T} \right)                          \nonumber \\
  & \equiv &   \omega_{\rm D} + \omega_{\rm F},
  \label{bp}
\end{eqnarray}
where 
\begin{equation}
  S_0 = \frac{1}{ \gamma^\mu p_\mu - M_t 
       -  {1 \over 2} \gamma^\mu q_\mu \gamma_5 \, \tau_3 }, \hskip0.6cm
  ( p^0 = i \, \omega_j + \mu  = i \, ( 2 j +1 ) \pi T + \mu  ).
  \label{bq}
\end{equation}
The propagator $S_0$ has four energy poles,
  $ p_0 = \epsilon_n ( n=1,2,3,4 )$, with
\begin{eqnarray}
  \epsilon_1 &=& \epsilon_- 
    =  \sqrt{ p_\perp^2 + \left( \, \sqrt{ M_t^2+p_z^2} - {q \over 2} \,  \right)^2 } >0,             \nonumber  \\
  \epsilon_2 &=& \epsilon_+ 
      =  \sqrt{ p_\perp^2 + \left( \, \sqrt{ M_t^2+p_z^2} +  {q \over 2}  \,  \right)^2 } >0,             \nonumber  \\
  \epsilon_3 &=& - \epsilon_1,         \nonumber  \\
  \epsilon_4 &=& - \epsilon_2.
  \label{br}
\end{eqnarray}
The $\omega_{\rm D}$ and $ \omega_{\rm F} $ have the form \cite{rf:TatNak,rf:NakTat},
\begin{eqnarray}
   \omega_{\rm D} &\equiv & -N_c  N_f   
  \int \frac{ d^3 p}{ (2 \pi)^3 } \left[ \, ( \epsilon_1 + \epsilon_2 ) 
    + T \ln \left( 1+ \exp \left\{ - \frac{ ( \epsilon_1 +\mu ) }{T} \right\} \right)
                                  \right.    \nonumber  \\
    & & \hskip5cm  \left.  + T \ln \left( 1+ \exp \left\{ - \frac{ ( \epsilon_2 +\mu ) }{T} \right\} \right)
                                           \right],      \label{bsa}            \\
  \omega_{\rm F} &\equiv & -N_c  N_f   
  \int \frac{ d^3 p}{ (2 \pi)^3 } \left[ \, 
     T \ln \left( 1+ \exp \left\{ - \frac{ ( \epsilon_1 - \mu ) }{T} \right\} \right)
                                  \right.    \nonumber  \\
    & & \hskip4cm  \left.  + T \ln \left( 1+ \exp \left\{ - \frac{ ( \epsilon_2 - \mu ) }{T} \right\} \right)
                                                                                                  \right].          
  \label{bsb}
\end{eqnarray}
The $\omega_{\rm D} $ represents the contribution from the Dirac sea
  and $\omega_{\rm F} $ the contribution from the Fermi sea.
The next leading order of $O(V_m \,)$ is \cite{rf:Kap}
\begin{eqnarray}
  & & -{T \over V}  \,  \frac{ \int {\cal D}  \, \bar\psi'  {\cal D} \psi'  A_m \exp ( A_0 ) }
                        { \int {\cal D}  \, \bar\psi'  {\cal D} \psi'          \exp ( A_0 ) }           \nonumber \\
  &=&  -{T \over V} \, N_c (-1) \, m \left\{ \frac{  \sin ( L \, q/2 )  }{  ( L \, q/2 )  } -1 \right\}
       V  \sum_{j=-\infty}^{\infty}  \int  \frac{ d^3 p}{ (2 \pi)^3 } \, {\rm tr} S_0          \nonumber \\
  & \equiv &   \delta  \omega_{\rm D} + \delta \omega_{\rm F},
  \label{bt}
\end{eqnarray}
where \cite{rf:Mae}
\begin{eqnarray}
   \delta  \omega_{\rm D} & \equiv & \left\{  \frac{  \sin ( L \, q/2 )  }{  ( L \, q/2 )  } -1 \right\} 
                  m \, N_c N_f  \int \frac{ d^3 p}{ (2 \pi)^3 }                                 \nonumber  \\
  & & \hskip0.9cm  \times
    \left[ \sum_{n=3,4} \frac{ M_t  \left\{ \sqrt{ M_t^2+p_z^2} + (-1)^n \left( q \over 2 \right)  \right\} } 
           { \epsilon_n \, \sqrt{ M_t^2+p_z^2} }  
            \left( 1+ \exp \left\{  \frac{ ( \epsilon_n - \mu ) }{T} \right\} \right)^{-1}      \right]    \nonumber  \\
  & = & \left\{  \frac{  \sin ( L \, q/2 )  }{  ( L \, q/2 )  } -1 \right\}
               m \, \frac{\partial}{\partial M_t} \, \omega_{\rm D}, 
     \label{bua}           
  \end{eqnarray}
\begin{eqnarray}
 \delta  \omega_{\rm F} & \equiv &   \left\{ \frac{  \sin ( L \, q/2 )  }{  ( L \, q/2 )  }  -1 \right\}     
                      m \, N_c N_f  \int \frac{ d^3 p}{ (2 \pi)^3 }                                       \nonumber  \\
  & & \hskip0.9cm  \times
    \left[ \sum_{n=1,2} \frac{ M_t  \left\{ \sqrt{ M_t^2+p_z^2} + (-1)^n \left( q \over 2 \right)  \right\} } 
           { \epsilon_n \, \sqrt{ M_t^2+p_z^2} }  
            \left( 1+ \exp \left\{  \frac{ ( \epsilon_n - \mu ) }{T} \right\} \right)^{-1}      \right]                      \nonumber  \\
  & = & \left\{  \frac{  \sin ( L \, q/2 )  }{  ( L \, q/2 )  } -1 \right\} 
                  m \, \frac{\partial}{\partial M_t} \, \omega_{\rm F}.
  \label{bub}
\end{eqnarray}
Both $ \delta \omega_{\rm D}$ and $\delta \omega_{\rm F}$  vanish in the
  chiral limit,
\begin{equation}
  \lim_{ m \rightarrow 0}  \delta \omega_{\rm D} 
  =  \lim_{ m \rightarrow 0}  \delta \omega_{\rm F} =0,
  \label{buc}
\end{equation}
as well as in the limit, $ q \rightarrow 0$, as it should be.

Eventually, thermodynamic potential $\omega$ up to $O(V_m)$ becomes
\begin{equation}
 \omega = \omega_{\rm D} + \omega_{\rm F} +{ (M_t -m)^2 \over 4G }
    + \delta  \omega_{\rm D} + \delta \omega_{\rm F} + O(V_m^2),
  \label{bv}
\end{equation}
where $\delta  \omega_{\rm D} $ and  $\delta  \omega_{\rm F} $ 
come from the term, $V_m$.
In the zero temperature limit, $ T \rightarrow 0$, each term becomes
\begin{eqnarray}
   \omega_{\rm D} & = & -N_c  N_f   
  \int \frac{ d^3 p}{ (2 \pi)^3 } \, ( \epsilon_1 + \epsilon_2 ),        \label{bwa}      \\
  \omega_{\rm F} & = & -N_c  N_f   
  \int \frac{ d^3 p}{ (2 \pi)^3 } \left[ \,  ( \mu - \epsilon_1 ) \, \theta ( \mu - \epsilon_1 ) 
       +  ( \mu - \epsilon_2 ) \, \theta ( \mu - \epsilon_2 )  \right],       \label{bwb}             \\
   \delta  \omega_{\rm D} & = &    
                  \left\{  \frac{  \sin ( L \, q/2 )  }{  ( L \, q/2 )  } -1 \right\}  \,
                     m \, \frac{\partial}{\partial M_t} \, \omega_{\rm D},                           \label{bwc}          \\
 \delta  \omega_{\rm F} & = &    \left\{  \frac{  \sin ( L \, q/2 )  }{  ( L \, q/2 )  } -1 \right\} 
                    m \, \frac{\partial}{\partial M_t} \, \omega_{\rm F}.    
  \label{bwd}
\end{eqnarray}
Note that $ \omega_{\rm F} $ is always negative, 
  $ \omega_{\rm F} \leq 0 $. 
But $ \delta \omega_{\rm F} $ can have either positive or negative value.
%
%
%
\section{Numerical calculation with zero temperature}

When the values of chemical potential $\mu$ and temperature $T$
  are given, the ground state of our system can be described by
  two parameters ($M_t, q$) which minimize the thermodynamic
  potential $\omega$.
We will obtain the ground state in the zero temperature case $T=0$ by
  numerical calculations.
Because the NJL model is a cut-off theory, a regularization method
  should be specified to define the theory, and we utilize the 
 proper-time  regularization method \cite{rf:Sch} here.
With the proper-time  regularization, $\omega_{\rm D}$ has the
  following form \cite{rf:TatNak,rf:NakTat},
\begin{eqnarray}
   \omega_{\rm D} & = & N_c  N_f  \, \frac{1}{ 4 \pi^{3/2} } 
         \int_{0}^{\infty} \frac{d k_z}{2 \pi}  \,
         \int_{1/ \Lambda^2}^{\infty} \frac{d \tau}{ \tau^{5/2} }  
        \left[ \exp \left\{ -  \left( \sqrt{ k_z^2 + M_t^2 } - {q \over 2} \, \right)^2  \tau  \right\} \right.
                                                       \nonumber  \\
   & & \hskip5.5cm  +  \left. \exp \left\{ -  \left( \sqrt{ k_z^2 + M_t^2 } + {q \over 2} \, \right)^2  \tau  \right\}
                                                       \right],
  \label{ca}
\end{eqnarray}
where $\Lambda$ is a cut-off parameter.
The regularized $\delta  \omega_{\rm D}$ can be obtained from
  Eq.(\ref{ca}) by use of Eq.(\ref{bua}).

We now set the three input parameters  $( G, m, \Lambda )$ 
  so that the observed values
  $ f_\pi =92.4 \, {\rm MeV} $ and $ m_\pi = 135 \, {\rm MeV} $
  are reproduced in the vacuum state $\mu =0$.
We first give arbitrary values of $\Lambda$, and then fix
  $G$ and $m$ so as to suit $f_\pi$ and $m_\pi$ to
 their observed values.
In Appendix A, the way of determining  $( G, m, \Lambda )$ is given
  in detail.
There still remains one degree of freedom, because we use
  only two observable quantities  $f_\pi$ and $m_\pi$ to
  determine the three input parameters.
Therefore, we require further that the chiral phase transition
  should be first order in spatially homogeneous case ($q=0$)
  at finite density.
The parameters  $( G, m, \Lambda )$ obtained in Appendix A are
  restricted by this requirement.
The spatially homogeneous case  ($q=0$) will be considered
  in the next subsection.
\subsection{Spatially homogeneous case at finite density }
Before studying spatially inhomogeneous case  ($q \neq 0$),
  we consider  in this subsection spatially homogeneous case  ($q = 0$),
  which is easier to deal with.
As is well known  \cite{rf:AsaYaz}, chiral symmetry is restored
  as the quark chemical potential $\mu$ increases (density increases).
Here we restrict the values  $( G, m, \Lambda )$ obtained in  Appendix A
  by requiring the chiral phase transition to be first order
  when $q=0$.
For a given value of $\mu$ (and $T=0$), the ground state is
  described by the parameter $M_t$ which minimize the
  thermodynamic potential $\omega$.
By the numerical calculations, we find that the chiral phase
  transition becomes first order when the input parameters
   $( G, m, \Lambda )$ obtained in  Appendix A satisfy the 
  following relation,
\begin{equation}
 G \Lambda^2 \geq 5.6.
  \label{cb}
\end{equation}
Henceforth, we use the input parameters' values 
   $( G, m, \Lambda )$ obtained in  Appendix A
which also satisfy the condition Eq.(\ref{cb}).

In analyzing the behavior of $M_t$, we find the following
  by numerical calculations.
As the chemical potential $\mu$ increases,
   the parameter $M_t$ decreases discontinuously
  and the first order phase transition of chiral symmetry occurs.
If $\mu$ increases further ($\mu \le \Lambda$), one might
  expect that the parameter $M_t$ will decrease more and
  approach $m$, $M_t > m$.
However, our observation is that, after chiral phase transition,
  $M_t$ decreases more and eventually it becomes 
  $ M_t < m$ when $\mu$ exceeds about $0.7 \Lambda$.
The relation  $ M_t < m$ implies that the dynamically obtained quark
  mass $M_t$ is less than the current quark mass if 
  $ \mu    \stackrel{>}{\sim}   0.7 \Lambda $, and this phenomenon
  is unnatural.
Hence we restrict ourselves to the region, 
  $ \mu    \stackrel{<}{\sim}   0.7 \Lambda $ and do not consider
  the value of $\mu$ larger than $  0.7 \Lambda $.
The fact that the relation $ M_t < m$ holds in the region
   $ \mu    \stackrel{>}{\sim}   0.7 \Lambda $ will be caused by our
  choice of the regularization method, that is the proper-time
  regularization.
This point will be discussed in Appendix B.

Before closing this subsection, we would like to study the effects of
   current mass on the spatially homogeneous chiral condensate.
This search would be helpful to investigate the effects of current
   mass on the spatially inhomogeneous chiral condensate
   which will be discussed in section 3.3.
In the ansatz, Eq.(\ref{bb}), when the chiral condensates are spatially
   homogeneous $(q=0)$, the condensates are chosen to be 
$ \langle \bar \psi \psi \rangle = C $ and
$ \langle {\bar \psi} i \gamma_5 \tau_3  \psi \rangle = 0$.
In the spatially homogeneous case, it is known that this choice
   realizes the lowest value of the energy if the current quark mass
   is finite.
Here, we make sure of this fact by the thermodynamic potential
   $\omega$ expanded in $V_m$. 
To this end, we assume in this subsection the following spatially homogeneous
   chiral condensate,
\begin{equation}
   \langle \bar\psi\psi  \rangle = C \cos \phi , \hskip 1cm
  \langle \bar\psi i \, \gamma_5 \, \tau_3 \, \psi \rangle = C \sin \phi,
    \hskip 1cm    (C<0),
  \label{cbb}
\end{equation}
where the chiral angle $\phi$ is a constant ( $ 0 \le \phi < 2 \pi $ ).
The thermodynamic potential having these condensates can be calculated
   as done in Section 2,
\begin{equation}
 \omega =  \left\{  \omega_{\rm D} + \omega_{\rm F} +{ M_t ^2 \over 4G }  \right\}
    + \left\{ -{m \over 2 G} M_t +  \delta  \omega_{\rm D} + \delta \omega_{\rm F} 
     +{m^2 \over 4 G}  \right\} + O(V_m^2),
  \label{cbc}
\end{equation}
where 
$ V_m  =  m \left\{  \exp \left( - i \gamma_5 \, \tau_3  \phi \right) -1 \right\}$.
Each term has the following form,
\begin{eqnarray}
   \omega_{\rm D} & = & -N_c  N_f   
  \int \frac{ d^3 p}{ (2 \pi)^3 } \,  2  \, \sqrt{ {\bf p}^2 + M_t^2 },                          \label{cbd}  \\
  \omega_{\rm F} & = & -N_c  N_f   
  \int \frac{ d^3 p}{ (2 \pi)^3 } \, 2 \,  \left( \mu -  \sqrt{ {\bf p}^2 + M_t^2 } \, \right) \, 
             \theta  \left( \mu -  \sqrt{ {\bf p}^2 + M_t^2 }  \right),                            \label{cbe}  \\
   \delta  \omega_{\rm D} & = &    
           ( \cos \phi -1 ) \, m (-1) N_c N_f 
             \int \frac{ d^3 p}{ (2 \pi)^3 } \,  2  \, 
              \frac{ M_t}{  \sqrt{ {\bf p}^2 + M_t^2 } }                                  \nonumber  \\
        &=&   ( \cos \phi -1 ) \, m \,  \frac{\partial}{\partial M_t} \, \omega_{\rm D},      \label{cbf}   \\
   \delta  \omega_{\rm F} & = &   
             ( \cos \phi -1 ) \, m \, N_c N_f 
             \int \frac{ d^3 p}{ (2 \pi)^3 } \,  2  \, 
              \frac{ M_t}{  \sqrt{ {\bf p}^2 + M_t^2 } } ~
                    \theta \left( \mu -  \sqrt{ {\bf p}^2 + M_t^2 }  \right)                               \nonumber  \\
        &=&   ( \cos \phi -1 ) \, m \,  \frac{\partial}{\partial M_t} \, \omega_{\rm F}.
  \label{cbg}
\end{eqnarray}
Thermodynamic potential $\omega$ is regarded as a function of two
   variables,  $ ( M_t, \phi )$, and the ground state is described by 
    $ ( M_t, \phi )$ which minimize $\omega$.
Since the $\omega$ depends on the chiral angle through $ \cos \phi$,
   one can constraint the chiral angle, $ 0 \le \phi \le \pi$.
The part, 
$  \left\{  \omega_{\rm D} + \omega_{\rm F} +{ M_t ^2 / 4G }  \right\} $,
   in Eq.(\ref{cbc}) does not depend on $\phi$.
The terms which represent the current mass effects are 
$   \left\{ - (m / 2 G ) M_t +  \delta  \omega_{\rm D} + \delta \omega_{\rm F}  \right\} $
   (the term, ${m^2 / 4 G} $, can be regarded as a constant because it does not
   depend on $M_t$ and $\phi$).
The term, $ - (m / 2 G ) M_t $, depends on only $M_t$, and this term
   lowers $\omega$.
Next, the term, $ \delta \omega_{\rm D} $, takes positive value,
$ \delta \omega_{\rm D} \ge 0 $.
This $ \delta \omega_{\rm D} $ is a monotone increasing function
   of the chiral angle $\phi$ in the range of $ 0 \le \phi \le \pi$,
   therefore the term,$ \delta \omega_{\rm D} $, tends to move
   the chiral angle $\phi$ closer to the value zero.
Thirdly, the term, $ \delta \omega_{\rm F} $, vanishes if 
   $ \mu \le M_t $, whereas it takes negative value 
   $ \delta \omega_{\rm F} <0 $ if $ \mu > M_t $.
This $ \delta \omega_{\rm F} $ is a monotone decreasing function of
   $\phi$  in the range of $ 0 \le \phi \le \pi$.
Therefore the term, $ \delta \omega_{\rm F} $, tends to move $\phi$
   closer to the value $\pi$, and this tendency rises when the chemical
   potential $\mu$ becomes larger.
We can say that the term $ \delta \omega_{\rm D} $ and the term
   $ \delta \omega_{\rm F} $ compete.
Then, how about their sum,
 $ ( \delta \omega_{\rm F} +  \delta \omega_{\rm D} )$ ?
The regularized $ \delta \omega_{\rm D} $ is obtained from Eq.(\ref{ca})
   by use of Eq.(\ref{cbf}), and the sum is
\begin{eqnarray}
   \delta \omega_{\rm F} +  \delta \omega_{\rm D}   
  &=& ( \cos \phi -1 ) \, 2 \, m \, N_c N_f  M_t \, (-1)                                  \nonumber \\
  &   \times &  \left[ { 1 \over 8 \pi^2 }  \int_{1/ \Lambda^2}^{\infty} d \tau {1 \over \tau^2} e^{- M_t^2 \tau } 
       -   \int \frac{ d^3 p}{ (2 \pi )^3 } \,
                 \frac{ 1}{  \sqrt{ {\bf p}^2 + M_t^2 } } ~
                    \theta \left( \mu -  \sqrt{ {\bf p}^2 + M_t^2 }  \right) \right].
  \label{cbp}
\end{eqnarray}
The equation enclosed in the brackets of the right-handed side is
   the same with that in the brackets in Eq.(\ref{pba}).
This equation takes positive value when $0 \le \mu <\mu_c $,
   and becomes zero if $ \mu = \mu_c $, as is discussed in Appendix B.
The equation is a monotone decreasing function of $\mu$.
Here, $\mu_c$ is defined in Eq.(\ref{pbb}) and takes the value,
   $ \mu_c \approx 0.7 \Lambda $ if $ M_t/ \Lambda \ll 1$.
The sum, $ (\delta \omega_{\rm F} +  \delta \omega_{\rm D} ) $,
   is then a monotone increasing function of $\phi$ in the range of
   $0 \le \mu <\mu_c $.
We therefore conclude that, in the range of $0 \le \mu <\mu_c $,
   $ (\delta \omega_{\rm F} +  \delta \omega_{\rm D} ) $
   tends to move the chiral angle $\phi$ closer to the value zero,
   and this tendency becomes weaker when $\mu$ grows.
From the above, we have shown that,  in the range of $0 \le \mu <\mu_c $,
   the condensate, 
$ \langle \bar \psi \psi \rangle =C, \langle {\bar \psi} i \gamma_5 \tau_3  \psi \rangle = 0,$
   realizes the lowest value of the thermodynamic potential
   $\omega$ if the chiral condensate is restricted to be spatially
   homogeneous.

The foregoing computation helps us discuss qualitatively
   the appropriateness of the ansatz Eq.(\ref{bb}) which we put
   for the spatially inhomogeneous chiral condensate.
When $\mu$ takes the value zero, the above computation concerning
   the spatially homogeneous condensate shows that the $\omega$
   does depend on the chiral angle $\phi$ and one cannot ignore
  that dependence.
Then Eq.(\ref{bb}) would not be appropriate if $\mu$ has small value because
   $ \langle \bar \psi \psi \rangle $ and
   $ \langle {\bar \psi} i \gamma_5 \tau_3  \psi \rangle $ 
   are put on the chiral circle in Eq.(\ref{bb}).
On the other hand, the dependence of $\omega$ on $\phi$ becomes
   weaker when the value of $\mu$ becomes larger.
Therefore we expect that
   the ansatz putting 
   $ \langle \bar \psi \psi \rangle $ and
   $ \langle {\bar \psi} i \gamma_5 \tau_3  \psi \rangle $ 
   on the chiral circle such as  Eq.(\ref{bb}) is a moderate assumption
   when $\mu$ takes large value if we neglect the contribution
   of order $O(V_m^2 \,)$.
%
%
%
\subsection{Numerical results}
As is discussed in the previous subsection, we restrict ourselves
  to the following region,
\begin{equation}
   G \Lambda^2 \geq 5.6,  \hskip1.5cm    \mu   \stackrel{<}{\sim}  0.7 \Lambda,
  \label{cc}
\end{equation}
the first condition comes from the requirement that the chiral
  phase transition is first order when $q=0$, and the second comes
  from the requirement that the dynamically obtained mass $M_t$
  is not less than the current quark mass $m$ when $q=0$.
The parameter $L$ is taken to be a enough large value,
  $ L = 2 \times 10^{8} / \Lambda $.
For a given value of $\mu$, the ground state is obtained by
  determining two values ($M_t, q$) which minimize the
  thermodynamic potential $\omega$.
If $q=0$ is the solution which minimizes $\omega$, the chiral
  condensate is spatially homogeneous.
%
%
%
%
\begin{figure}
  \begin{center}
     \includegraphics[height=7cm]{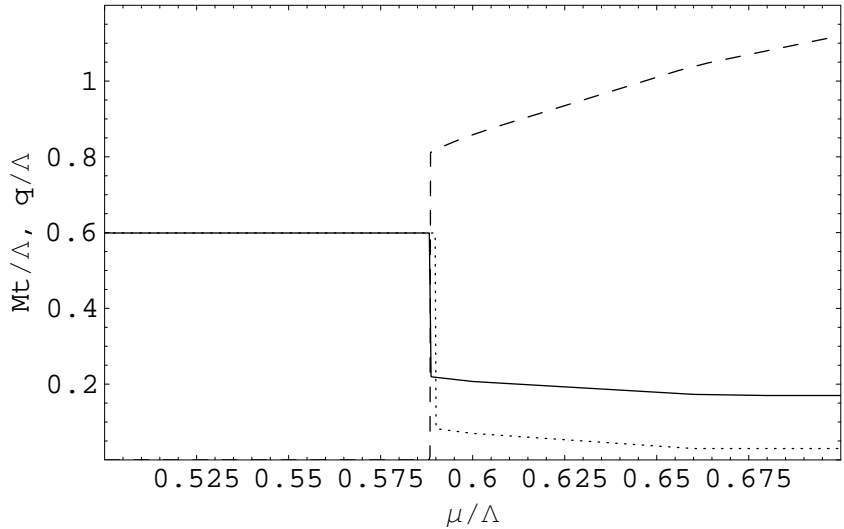}
  \end{center}
   \caption{The mass $M_t$ (solid) and the wave number $q$ (long-dashed) as a function
                  of the chemical potential $\mu$, in units of $\Lambda$. 
                   For comparison,  $M_t ( q=0 )$ without inhomogeneous chiral condensate is
                  also represented
                  by dotted line.}
\label{fig:1}
\end{figure}
On the other hand, if $q \neq 0$ is the solution which minimizes
   $\omega$, the chiral condensate is  spatially inhomogeneous.
According to numerical calculations, we always have the solution
  $q=0$ for arbitrary values of $\mu$ when
  $ 5.6 \leq G \Lambda^2 < 6.94 $.
Namely, in the region  $ 5.6 \leq G \Lambda^2 < 6.94 $,
  the ground state is not spatially inhomogeneous
  chiral condensate.

If $  G \Lambda^2 $ is larger than $6.94$, $ G \Lambda^2 \geq 6.94 $,
  however, there exist the solutions $q \neq 0$ for large values 
  of $\mu$.
Here we show a concrete example of numerical results with
  the input parameters,
  $  G \Lambda^2 =7.513,~ \Lambda = 635 {\rm MeV} $, and
  $ m= 15.3  {\rm MeV}, ( M_t \, ( \mu=0 ) = 380 {\rm MeV} ) $ 
  in Figure.1.
In the low density region, $ 0 \leq \mu < 0.5885 \Lambda $,
  we have $q=0$ and the chiral condensate is spatially homogeneous
  as usual.
In the high density region, $ 0.5885 \Lambda \leq \mu \leq 0.7 \Lambda $,
  we get $q \neq 0$ (long-dashed line) and spatially inhomogeneous
  dual chiral condensate is realized.
Let us see the behavior of $M_t$ represented by solid line.
Since $M_t$ decreases discontinuously at $ \mu = 0.5885 \Lambda $,
  the chiral phase transition is first order.
In this high density region, $ \mu \geq 0.5885 \Lambda $,
  one finds by numerical calculations that the relation,
  $  \mu < M_t+q/2 $, holds.
Therefore, the $\epsilon_2$ state is occupied with no quark
   because of $ \epsilon_2 \geq  M_t+q/2 $, and all quarks 
  in the Fermi sea are filled in the $\epsilon_1$ states.
For comparison, we also show a local minimum solution $ M_t ( q=0 )$
  with $q=0$, which is represented by dotted line in Figure.1.
In the region, $\mu \geq 0.5885 \Lambda $, the points
   $( M_t ( {\rm solid \, line} ), q \neq 0 ( {\rm long-dashed \, line} ) )$
   in the $M_t-q$ plane 
  correspond to the absolute minimum energy state (ground state)
  of $\omega$, and the points 
   $( M_t ( {\rm dotted \, line} ), q=0 )$
   correspond to the local minimum energy state.
The spatially inhomogeneous ($q \neq 0$)
  dual chiral condensate occurs at densities higher than about the
  usual ($q=0$) chiral phase transition point, $\mu=0.59 \Lambda $.
The characteristic feature in Figure.1 is as follows.
As the chemical potential $\mu$ increases over the value of
   $ 0.5885 \Lambda $, $M_t$ decreases a little.
Even at high density of $\mu=0.7 \Lambda $, the value of
  $M_t$ remains rather large, $ M_t \approx 0.17 \Lambda $.
This feature is different from the result obtained in the
  chiral limit case \cite{rf:TatNak,rf:NakTat}.
If $ G \Lambda^2 $ takes the values,
  $ 7.04  \leq  G \Lambda^2   \leq  8.1 $,
 the behavior of $q$ ($M_t$) in the $ q - \mu $ plane 
 (in the $ M_t - \mu $ plane) is similar to that in Figure.1.
On the other hand, if  $ G \Lambda^2 $ takes the values,
  $ 6.94  \leq  G \Lambda^2  <  7.04 $,
  the wave number $q$ remains zero at the value of chemical 
  potential $\mu_{\rm chi}$ at which $M_t$ decreases discontinuously.
The wave number $q$ becomes non-zero at the value of the chemical
  potential whose magnitude is over $\mu_{\rm chi}$.
\subsection{Effects of current mass}
In this subsection, the effect of current mass on the
  thermodynamic potential $\omega$, Eq.(\ref{bv}), 
  with $T=0$ is cleared.
We divide $\omega$ into two parts,
\begin{equation}
  \omega 
    =  \left\{ \omega_{\rm D} + \omega_{\rm F} +{ M_t^2 \over 4G } \right\}
       + \left\{
              -\frac{m}{2 G} \, M_t +(  \delta  \omega_{\rm D} + \delta \omega_{\rm F} )
               + \frac{m^2}{4 G} \right\} + O(V_m^2 \,).
  \label{ce}
\end{equation}
The thermodynamic potential $\omega$ is regarded as a function of the 
   two parameters $(M_t, q)$.
The terms $ \delta  \omega_{\rm D}$ and $  \delta \omega_{\rm F} $ are defined in 
   Eq.(\ref{bua}) and (\ref{bub}), and the effects of current mass $m$
   is described by the second brackets of Eq.(\ref{ce}).
Because the ground state is described by the two parameters
   $( M_t, q )$ which minimize $\omega$, the term $m^2 / 4 G$ 
   in the second brackets of Eq.(\ref{ce}) can be regarded as a constant.
We then need to see remaining two terms,
  $ -(m / 2G \,) M_t $ and 
  $ (  \delta  \omega_{\rm D} +\delta \omega_{\rm F} )$.
First, the term $ -(m / 2G \,) M_t $ in the second brackets of Eq.(\ref{ce}) depends on
  $M_t$, not on $q$.
The larger $M_t$ is, the more markedly this term 
  $ -(m / 2G \,) M_t $ lowers $\omega$ independently of $q$.
Therefore,  the term $ -(m / 2G \,) M_t $ has possibility 
  to make $\omega$ have minimum value at the point $q \neq 0$.

Second, let us inspect the term 
  $ (  \delta  \omega_{\rm D} +\delta \omega_{\rm F} )$.
We want to judge whether   $ \delta  \omega_{\rm D} $ and
  $ \delta \omega_{\rm F} $ tend to make $\omega$ minimized 
  at the point $q \neq 0$ or not.
To this end, the dependence of  
  $ \delta  \omega_{\rm D} $ ($ \delta \omega_{\rm F} $) on
  $q$ is found by expanding 
    $ \delta  \omega_{\rm D} $ ($ \delta \omega_{\rm F} $)
  about $q=0$,
\begin{equation}
   \delta  \omega_{\rm D} 
     = { 1 \over 2 !} \left.   \frac{ \partial^2 (  \delta  \omega_{\rm D} ) }{ \partial q^2 }
          \right |_{q=0} \cdot q^2 + O ( q^3 ).
  \label{cf}
\end{equation}
Since the coefficient of $q^2$,
\begin{equation}
    { 1 \over 2 !} \left.  \frac{ \partial^2 (  \delta  \omega_{\rm D} ) }
      { \partial q^2 } \right|_{q=0} 
   = {1 \over 6} \left( { L \over 2 } \right)^2 m N_c N_f \frac{M_t}{4 \pi^2}
         \int_{1/ \Lambda^2}^{\infty} d \tau {1 \over \tau^2} e^{- M_t^2 \tau }  ~ > 0,
  \label{cg}
\end{equation}
is positive, the value of $ \delta  \omega_{\rm D}$ at $q \neq 0$
  is larger than that at $q=0$ when $q$ is small.
Hence,  $ \delta  \omega_{\rm D}$ tends to make $\omega$
  not minimized at the point $q \neq 0$ if $q$ is small.
Similarly,  $ \delta  \omega_{\rm F}$ is expanded about $q=0$,
\begin{equation}
   \delta  \omega_{\rm F} 
     = { 1 \over 2 !} \left.  \frac{ \partial^2 (  \delta  \omega_{\rm F} ) }{ \partial q^2 }
          \right|_{q=0} \cdot q^2 + O ( q^3 ).
  \label{ch}
\end{equation}
Since the coefficient of $q^2$,
\begin{equation}
     { 1 \over 2 !} \left.  \frac{ \partial^2 (  \delta  \omega_{\rm F} ) }
        { \partial q^2 } \right|_{q=0} 
   = - {1 \over 6} \left( { L \over 2 } \right)^2 m N_c N_f \cdot   2 M_t
      \int \frac{ d^3 p}{ (2 \pi )^3 }
       \, {1 \over \epsilon} \, \theta ( \mu -\epsilon )  ~ < 0,
  \label{ci}
\end{equation}
is negative, the value of $ \delta  \omega_{\rm F}$ at $q \neq 0$
  is smaller than that at $q=0$ when $q$ is small.
Hence,  $ \delta  \omega_{\rm F}$ tends to make $\omega$
  minimized at the point $q \neq 0$ if $q$ is small.
Now, how about the sum, 
  $  ( \delta  \omega_{\rm D} +  \delta  \omega_{\rm F} )$ ?
Expanding  $  ( \delta  \omega_{\rm D} +  \delta  \omega_{\rm F} )$ 
  about $q=0$, we obtain
\begin{equation}
   \delta  \omega_{\rm D} +\delta  \omega_{\rm F} 
     = { 1 \over 2 !} 
          \left.  \frac{ \partial^2 (  \delta  \omega_{\rm D} 
         + \delta  \omega_{\rm F} )  }{ \partial q^2 }
          \right|_{q=0} \cdot q^2 + O ( q^3 ),
  \label{cj}
\end{equation}
where the coefficient of $q^2$ is
\begin{equation}
   {1 \over 6} \left( { L \over 2 } \right)^2 2 \, m N_c N_f  M_t
       \left[ { 1 \over 8 \pi^2 }  \int_{1/ \Lambda^2}^{\infty} d \tau
       {1 \over \tau^2} e^{- M_t^2 \tau } 
       -   \int \frac{ d^3 p}{ (2 \pi )^3 }
               \, {1 \over \epsilon} \, \theta ( \mu -\epsilon )  \right].
  \label{ck}
\end{equation}
The equation enclosed in the above brackets is the same with the
  equation in the brackets in Eq.(\ref{pba}), which equation takes
  positive value when $ \mu \leq \mu_c ( \approx 0.7 \Lambda ) $,
  as is discussed in Appendix B.
The coefficient Eq.(\ref{ck}) is positive if 
   $ \mu \leq \mu_c ( \approx 0.7 \Lambda ) $.
We can say that when $q$ is small the value of 
   $  ( \delta  \omega_{\rm D} +  \delta  \omega_{\rm F} )$ 
  at $q \neq 0$ is larger than that at $q=0$.
Furthermore, we confirm by numerical calculations that 
  $  ( \delta  \omega_{\rm D} +  \delta  \omega_{\rm F} )$ 
  takes positive value in the range of $ 0 \leq  q  \leq  2 \mu $
  when  $ \mu \leq   0.7 \Lambda $.
Eventually, we find that 
   $  ( \delta  \omega_{\rm D} +  \delta  \omega_{\rm F} )$ 
  tends to make $\omega$ not minimized at the point $q \neq 0$.
In other words, 
   $  ( \delta  \omega_{\rm D} +  \delta  \omega_{\rm F} )$ 
  tends not to
realize the inhomogeneous chiral condensate ($q \neq 0$).

Our argument in this subsection is summarized as follows.
The influence of the current mass $m$ on the thermodynamic
  potential $\omega$ is represented as 
  $ -(m / 2G \,) M_t 
   + (  \delta  \omega_{\rm D} + \delta \omega_{\rm F} )$.
The term $ -(m / 2G \,) M_t $ tends to make $\omega$ minimized
  at the point $q \neq 0$.
On the other hand, the term
  $  ( \delta  \omega_{\rm D} +  \delta  \omega_{\rm F} )$ 
  tends to make $\omega$ not
minimized at the point $q \neq 0$.
Whether the introduction of the finite current mass tends to
  materialize the inhomogeneous chiral condensate ($q \neq 0$)
or not depends on the competition between the term
  $ -(m / 2G \,) M_t $ and the term
   $  ( \delta  \omega_{\rm D} +  \delta  \omega_{\rm F} )$.
\section{Spatially inhomogeneous baryon density}
In the previous subsection, we consider the effect of the
  finite current mass on the ground state with the spatially
  inhomogeneous chiral condensate.
The influence of the current mass on the thermodynamic potential
  is represented as 
  $ -(m / 2G \,) M_t 
   + (  \delta  \omega_{\rm D} + \delta \omega_{\rm F} )$.
There is, however, another influence which the current
  mass has on the system.
In the chiral limit, the baryon density is spatially homogeneous 
  even if the spatially inhomogeneous chiral condensate 
  occurs \cite{rf:TatNak,rf:NakTat}.
In contrast, when the current mass is finite, the baryon density
  becomes spatially inhomogeneous if the spatially inhomogeneous
  chiral condensate occurs.
This will be shown as follows.

The quark number (one-thirds of the baryon number) density
  $\rho$ can be written by the field $ \psi' $ defined in 
  section 2,
\begin{eqnarray}
  \rho &=& \langle  \psi^\dagger  \psi \rangle                                                          \nonumber    \\
    &=&  \langle  \psi'^\dagger  \psi' \, \rangle                                                          \nonumber    \\
    &=& -i \, {\rm tr} \left[ \gamma^0 S ( x, x ) \right]                                   \nonumber    \\
    &=&  -i \, {\rm tr} \left[ \gamma^0 S_0 ( x, x ) \right] 
            -i \,  \int d^4 y {\rm tr} \left[ \gamma^0 S_0 ( x, y ) V_m (y) \, S_0 (y, x ) \right] + O (V_m^2 )               \nonumber    \\
    &=& -i \, \left( i T \sum_{j=-\infty}^{\infty} \int \frac{d^3 p}{ (2 \pi )^3 } \right) {\rm tr} \left[ \gamma^0 S_0 ( p) \right]   \nonumber    \\
    & & \hskip0.5cm  -i m \left\{
           - \left( i T \sum_{j=-\infty}^{\infty}  \int \frac{d^3 p}{ (2 \pi )^3 } \right)  
            {\rm tr} \left[ \gamma^0 S_0 ( p)  S_0 ( p) \right]        \right.                          \nonumber    \\
    & &  \hskip1.7cm  + \exp ( -i \mbox{\boldmath  $q$} \cdot \mbox{\boldmath  $r$} ) \left( i T
\sum_{j=-\infty}^{\infty}  \int
\frac{d^3 p}{ (2 \pi )^3 } \right) 
             {\rm tr} \left[ \gamma^0 S_0 ( p)  \frac{ ( 1+ \gamma_5 \tau_3 ) }{2}  S_0 ( p+q) \right]                \nonumber    \\
    & &  \hskip1.7cm  \left.  + \exp ( i \mbox{\boldmath  $q$} \cdot \mbox{\boldmath  $r$} ) 
\left( i T \sum_{j=-\infty}^{\infty}  \int
\frac{d^3 p}{ (2 \pi )^3 } \right) 
             {\rm tr} \left[ \gamma^0 S_0 ( p)  \frac{ ( 1- \gamma_5 \tau_3 ) }{2}  S_0 ( p-q) \right]   
                                                              \right\}                            \nonumber    \\
   & &  \hskip0.5cm + O ( V_m^2 ),
  \label{da}
\end{eqnarray}
where $S$ is the propagator of $ \psi' $, and the perturbative
  approximation Eq.(\ref{bk}) has been used.
Since $\rho$ depends on 
  $  \mbox{\boldmath  $r$} $,
  the baryon number density is spatially inhomogeneous.
The wave number representing inhomogeneousness of 
  baryon number density coincides with that of chiral condensate
  up to the order $O(V_m)$.
In the chiral limit, the baryon number density becomes spatially
  homogeneous, indeed.
Why does the baryon number density become spatially
  inhomogeneous when the current mass is finite?
The path integral representation Eq.(\ref{be}) will help us
  to see this.
The propagator of the field  $ \psi $ depends on the space
  coordinates $\mbox{\boldmath  $r$}$ explicitly.
In the chiral limit, however, the propagator $S$ of the field
   $ \psi' $ introduced in Eq.(\ref{be}) does not depend on 
  $\mbox{\boldmath  $r$}$.
One can write $\rho$ in terms of the propagator $S$ in the
  mean-field approximation.
On the other hand, when the current mass is finite, the propagator
  $S$ does depend on  $\mbox{\boldmath  $r$}$ explicitly.
\section{Conclusion}
We have studied how the introduction of finite current quark mass
  affects the ground state with the spatially inhomogeneous
  chiral condensate, Eq.(\ref{bb}), which is actually realized in
  the NJL model at finite density in the chiral limit.
Our numerical calculations show that, even if the finite current
  quark mass is introduced, the spatially inhomogeneous
  chiral condensate can take place.
If $G \Lambda^2 \geq 6.94 $ is satisfied, the spatially inhomogeneous
  chiral condensate occurs (i.e., $q \neq 0$) at high density.
When $ 5.6 \leq G \Lambda^2  < 6.94 $, the spatially inhomogeneous
  chiral condensate does not occur (i.e., $q = 0$).
The ground state was determined by obtaining the values of
  two parameters ($M_t, q$) which minimize the thermodynamic
   potential $\omega$.
When the current mass $m$ takes a finite small value, the 
  correction due to $m$ was added to $\omega$.
That correction was divided into the following two parts.
The first part is the term, $ -(m / 2G \,) M_t $, which depends
  on the constituent mass $M_t$ but not on the wave number $q$.
This term lowers $\omega$ more markedly for the larger $M_t$,
  independently of $q$.
Hence, this first part, $ -(m / 2G \,) M_t $, has possibility to tend
  to realize the spatially inhomogeneous chiral condensate.
The second part is the term written as 
  $ ( \delta  \omega_{\rm D} +  \delta  \omega_{\rm F} )$,
  which depends on both $M_t$ and $q$.
According to numerical calculations, the value of 
   $ ( \delta  \omega_{\rm D} +  \delta  \omega_{\rm F} )$
  at $q \neq 0$ is larger than that at $q=0$ for each
  value of $M_t$.
Therefore, the second part,
   $ ( \delta  \omega_{\rm D} +  \delta  \omega_{\rm F} )$
  tends not to realize the spatially inhomogeneous
  chiral condensate.
Whether the introduction of the finite current mass tends
  to realize the spatially inhomogeneous chiral condensate 
  or not depends on the competition between the term
   $ -(m / 2G \,) M_t $ and the term
   $ ( \delta  \omega_{\rm D} +  \delta  \omega_{\rm F} )$.

The introduction of the current mass influences not only
  the possibility of the spatially inhomogeneous chiral condensate 
  but also space dependence of the baryon number density.
In the chiral limit, the baryon number density is spatially homogeneous
  even if the spatially inhomogeneous chiral condensate 
  occurs \cite{rf:TatNak,rf:NakTat}.
In contrast, when the current mass takes a finite value, the
  baryon number density becomes spatially inhomogeneous
  if the spatially inhomogeneous chiral condensate is realized.
In addition, the wave number vector of the spatially inhomogeneous
  baryon number density coincides with that of the spatially
  inhomogeneous chiral condensate
  when the value of the current mass is small.

Here we give consideration to the ansatz Eq.(\ref{bb}) that describes
  the spatially inhomogeneous dual chiral condensate.
In the chiral limit, the Lagrangian is chiral invariant and this
  ansatz seems to be appropriate because 
  $  \langle \bar\psi\psi  \rangle $ and
  $  \langle \bar\psi  i \, \gamma_5 \, \tau_3 \,  \psi  \rangle $
  are put on a chiral circle,
  $ \langle \bar\psi\psi \rangle^2 + \langle \bar\psi i \, \gamma_5 \,
     \tau_3 \, \psi \rangle^2  = C^2 $.
On the other hand, when the current mass is finite, 
   the Lagrangian is not chiral invariant and there might exists more
  appropriate ansatz than Eq.(\ref{bb}).
In such ansatz, 
  $  \langle \bar\psi\psi  \rangle $ and
  $  \langle \bar\psi  i \, \gamma_5 \, \tau_3 \,  \psi  \rangle $
  are not put on the chiral circle and would satisfy more
  complicated relation.
In the present paper dealing with the finite current mass, however,
  we have assumed the ansatz Eq.(\ref{bb}) for simplicity.
Within the limits of this ansatz, we found the ground state with
  $M_t >0$ and $q \neq 0$.
Also, the baryon number density was found to be spatially
  inhomogeneous as well as the chiral condensate.
If we can find more appropriate ansatz than Eq.(\ref{bb}),
  we would be able to have thermodynamic potential
  whose minimum value is smaller than the minimum value
  obtained by use of the ansatz Eq.(\ref{bb}).
With such more appropriate ansatz than  Eq.(\ref{bb}), physical quantities such as
   the baryon number density will be affected and the way of the phase
   transition might be influenced.
Besides our ansatz, other type of ansatz was studied in Ref \cite{rf:Nic1,rf:Nic2} as discussed
   in section 1.
In order to determine which ansatz is to be realized, one needs to compare
   the thermodynamic potentials between them.

Several problems are still remained.
First, it will be interesting to carry out numerical calculations
  with finite temperature $T>0$.
The phase diagram in the $T-\mu$ plane should be affected when
  the current quark mass
  is taken to be non-zero \cite{rf:Nic1,rf:Nic2}.
Second, the study with the inclusion of color superconductivity.
In the chiral limit, the phase diagram in  the $T-\mu$ plane 
  including a second color superconducting phase (2SC),
  uniform chiral and non-uniform chiral phase is
  studied \cite{rf:Sad}.
The introduction of non-zero current quark mass should
  influence the diagram.
Finally, 
   in the NJL model with quark mass term
  there appears the Nambu-Goldstone boson (NG boson)
  when the ground state with spatially inhomogeneous
  chiral condensate is realized.
Since this inhomogeneous chiral condensate breaks space
  translational symmetry spontaneously,
  the NG boson appears \cite{rf:CasNar}.
In the low energy phenomena, the NG boson would play
  an important role.
%
%
\section*{Acknowledgments}

The author thanks the Yukawa Institute for Theoretical Physics at Kyoto University.
Discussions during the YITP workshop YITP-W-08-09
   on "Thermal Quantum Field Theories and Their Applications" were useful to 
   complete this work.
%
%
%
%
%
%
\newpage 
\noindent{\Large\bf Appendix}
\appendix 
%
%
\section{Input parameters}
\renewcommand{\theequation}{A.\arabic{equation}}
\setcounter{equation}{0}
We explain in this appendix a way to set the values of the
  input parameters  $( G, m, \Lambda )$ of the NJL model
  in the proper-time regularization (PTR) scheme.
In the vacuum state $\mu=0$, the pion decay constant
  $f_\pi$ in the PTR scheme is represented as \cite{rf:Kle}
\begin{equation}
   f_\pi^2 
   = \frac{ N_c \, M_t^2 }{4 \pi^2 } \int_{ M_t^2 / \Lambda^2 }^{\infty}
   d t \, { 1 \over t} \, e^{-t},
  \label{paa}
\end{equation}
and the pion mass $m_\pi$ satisfies the following relation \cite{rf:Kle},
\begin{equation}
  m_\pi^2 \, f_\pi^2 
   =  m \, {1 \over 2 G} \, M_t.
  \label{pab}
\end{equation}
The gap equation in the PTR scheme becomes  \cite{rf:Kle}
\begin{equation}
  M_t 
  =  m + {1 \over 2 \pi^2} G N_c N_f M_t \, F_2 ( M_t^2,  { 1 \over \Lambda^2 } ),
  \label{pac}
\end{equation}
where the function $F_2$ is defined as
\begin{equation}
   F_2 ( M_t^2,  { 1 \over \Lambda^2 } )
  \equiv    \int_{ 1 / \Lambda^2 }^{\infty}   d s \, { 1 \over s^2} \, e^{- M_t^2 s}
  =  \Lambda^2 \int_{ M_t^2 / \Lambda^2 }^{\infty}
              d t \, { 1 \over t^2} \, e^{-t}.
  \label{pad}
\end{equation}
Now, we give an arbitrary value to $\Lambda$ firstly.
Then, $m$ and $G$ are determined so as to reproduce the values
   $ f_\pi =92.4 \, {\rm MeV} $ and $ m_\pi = 135 \, {\rm MeV} $.
When the value of $\Lambda$ is given, one can calculate the value
  of $M_t$ through Eq.(\ref{paa}).
Regarding the equations Eq.(\ref{pab}) and Eq.(\ref{pac}) 
  as simultaneous equations for $G$ and $m$, one can obtain
  both the values of  $G$ and $m$.
Thus, we can determine the values of  $G$ and $m$ for
  a given value of $\Lambda$.
%
%
%
%
\section{Chiral symmetry restoration and \\
   proper-time regularization}
\renewcommand{\theequation}{B.\arabic{equation}}
\setcounter{equation}{0}
In this appendix, we look into the reason why the $M_t$
  becomes less than the current mass, $ M_t < m$ for
  $q=0$ and large chemical potential 
  $ \mu  \stackrel{>}{\sim}  0.7 \Lambda $
  in the numerical calculations.
At the minimum point of the thermodynamic potential
  $\omega$ with $q=0$ and $T=0$, the condition,
  $ \delta \omega / \delta M_t =0$, is held,
\begin{equation}
  M_t - m = 4 G N_c N_f M_t \left[ \frac{1}{8 \pi^2 } \int_{1/\Lambda^2}^{\infty} 
      d s {1 \over s^2} \exp ( -M_t^2 s ) - \int \frac{ d^3 p}{ (2 \pi )^3 }
       \, {1 \over \epsilon} \, \theta ( \mu-\epsilon ) \right],
  \label{pba}
\end{equation}
where $ \epsilon = \sqrt{ \mbox{\boldmath  $p$}^2 + M_t^2 }$.
This is nothing but the gap equation and it should be noted
  that the PTR scheme has been used in this equation.
The first (second) term in the brackets of Eq.(\ref{pba}) 
  represents the contribution from the Dirac sea (Fermi sea).
The second term increases if the chemical potential $\mu$
  becomes larger.
When the chemical potential reaches a certain value $\mu_c$,
  the second term becomes equal to the first term,
\begin{equation}
    \frac{1}{8 \pi^2 } \int_{1/\Lambda^2}^{\infty} 
      d s {1 \over s^2} \exp ( -M_t^2 s )  =  \int \frac{ d^3 p}{ (2 \pi )^3 }
       \, {1 \over \epsilon} \, \theta ( \mu_c -\epsilon ), 
  \label{pbb}
\end{equation}
and then one has $M_t=m$ from Eq.(\ref{pba}).
If the chemical potential $\mu$ is larger than $\mu_c$,
  the second term becomes larger than the first term and
  one has $M_t < m$.
In short the relationship between $M_t$ and $m$ in terms of
  size is summarized as follows.
For small values of the chemical potential $\mu$, $M_t$ is
larger than $m$.
When $\mu$ reaches $\mu_c$, we have $M_t=m$.
If $\mu$ becomes larger than  $\mu_c$, we have $M_t < m$.
Now, let us estimate the value of  $\mu_c$ assuming
  $ M_t/ \Lambda \ll 1 $.
The expansion of the left side of Eq.(\ref{pbb}) about
   $ M_t/ \Lambda =0 $ takes the form \cite{rf:Kle}
\begin{eqnarray}
    \frac{1}{8 \pi^2 } \int_{1/\Lambda^2}^{\infty} d s {1 \over s^2} \exp ( -M_t^2 s )  
    &=&   \frac{\Lambda^2}{8 \pi^2 } \left( \frac{ M_t^2}{ \Lambda^2 } \right)
                     \int_{M_t^2 / \Lambda^2}^{\infty} d t {1 \over t^2} \exp ( -t )     \nonumber  \\
   &=&  \frac{\Lambda^2}{8 \pi^2 } \left[ 1 + ( \gamma -1 ) 
                \left( \frac{ M_t^2}{ \Lambda^2 } \right)  + \cdots  \right].
 \label{pbc}
\end{eqnarray}
The expansion of the right side of Eq.(\ref{pbb}) about
   $ M_t/ \Lambda =0 $ is
\begin{equation}
 \frac{\Lambda^2}{4 \pi^2 } \left[  \left( \frac{ \mu_c^2}{ \Lambda^2 } \right)    
               -{1 \over 2}  \left( \frac{ M_t^2}{ \Lambda^2 } \right)  + \cdots  \right].
  \label{pbd}
\end{equation}
Neglecting the order $ O ( M_t^2 / \Lambda^2 )$ in Eq.(\ref{pbb}),
  we have
\begin{equation}
  \mu_c \approx \frac{ \Lambda}{ \sqrt{2} } \approx 0.7 \Lambda.
  \label{pbe}
\end{equation}
Thus, we can understand roughly why $M_t$ becomes less
  than $m$ for $ \mu  \stackrel{>}{\sim}  0.7 \Lambda $ 
  in the PTR scheme.

It would be helpful to examine the relationship between $M_t$
  and $m$ in a different regularization scheme, say the 
  three-momentum cut-off scheme  \cite{rf:Kle}.
In this scheme, the first term in the brackets of the gap
  equation Eq.(\ref{pba}) is replaced by
\begin{equation}
   \int \frac{ d^3 p}{ (2 \pi )^3 } \, {1 \over \epsilon} \, 
    \theta  \, ( \, \Lambda_{\rm 3-momentum} - \vert  \mbox{\boldmath  $p$} \vert \, ). 
  \label{pbf}
\end{equation}
The value of the chemical potential at which the equation 
  enclosed in brackets of the gap equation becomes zero is
\begin{equation}
   ( \mu_c )_{\rm 3-momentum} \approx  \Lambda_{\rm 3-momentum}.
  \label{pbg}
\end{equation}
Therefore, $M_t$ does not become less than $m$ for 
  $ \mu  \stackrel{<}{\sim}  \Lambda_{\rm 3-momentum} $ 
  in the three-momentum cut-off scheme.
\vskip 2cm
%
%
%
%
%
%
%
%

%
%
%
\end{document}